# Let's Resonate! How to Elicit Improvisation and Letting Go in Interactive Digital Art

Jean-François Jégo
INREV team, AIAC Laboratory
University Paris 8
Saint-Denis, France
jean-francois.jego03@univ-paris8.fr

Margherita Bergamo Meneghini
Choreographer
Compagnie Voix
Chalon-sur-Saône, France
margherita@compagnievoix.com

## ABSTRACT

Participatory art allows for the spectator to be a participant or a viewer able to engage actively with interactive art. Real-time technologies offer new ways to create participative artworks. We hereby investigate how to engage participation through movement in interactive digital art, and what this engagement can awaken, focusing on the ways to elicit improvisation and letting go. We analyze two Virtual Reality installations, "InterACTE" and "Eve, dance is an unplaceable place," involving body movement, dance, creativity and the presence of an observing audience. We evaluate the premises, the setup, and the feedback of the spectators in the two installations. We propose a model following three different perspectives of resonance: 1. Inter Resonance between Spectator and Artwork, which involves curiosity, imitation, playfulness and improvisation. 2. Inner Resonance of Spectator him/herself, where embodiment and creativity contribute to the sense of being present and letting go. 3. Collective Resonance between Spectator/Artwork and Audience, which is stimulated by curiosity, and triggers motor contagion, engagement and gathering. The two analyzed examples seek to awaken open-minded communicative possibilities through the use of interactive digital artworks. Moreover, the need to recognize and develop the idea of resonance becomes increasingly important in this time of urgency to communicate, understand and support collectivity.

## CCS CONCEPTS

• **Applied computing** → Arts and humanities → Performing arts, → Media arts • **Human-centered computing** → Collaborative and social computing → Empirical studies in collaborative and social computing

## KEYWORDS

Embodied cognition, resonance, movement, interactive art, spectator, improvisation, dance and technology, virtual reality, philosophical perspectives

## 1. CONTEXT AND DEFINITIONS

### 1.1 Motivations

From the early sixties, the emergence of participatory art has allowed the audience to become participants in the event and thus to engage in the creative process. More recently, interactive technologies have offered new ways to create participative artworks and performance. In this article, we consider the spectator as a participant or a viewer who is free to engage, or not, with interactive art and performance [12]. Starting from this premise, we propose to investigate ways to engage participation through movement in interactive digital art, and the possibilities this participation can awaken. More precisely, we wish to focus on how to elicit improvisation and letting go in spectators.

### 1.2 Movement and aesthetic experience

Thoughts about the role of movement (and by extension human body movement) in understanding and experiencing reality have affected man for centuries, taking many different pathways but always existing in some form. In general terms, for Heraclitus [23] in the 6th century BC, movement was conceived as the essential foundation of reality. Subsequently, Spinoza [14] in the 17th century AD imbued the body and its movement with the implicit and unique experience of each person, his/her incarnation in one reality. More recently in our contemporaneity, Deleuze [11], citing Nietzsche and Kierkegaard, spoke of the "production of movement to move the spirit out of any representation [...] to invent vibrations, rotations, vortices, gravitation, dances and jumps that touch directly the spirit."
Couchot et al. [9] noted that during the aesthetic experience when a spectator attends a "jam session" or a theater/street play, the actors who improvise invite the spectators to enter directly in motor and emotional empathy with them. The authors explain that empathy has two basic components: the first is a motor and emotional resonance triggered automatically, and the second a controlled and intentional subjective perspective of the other. They also make the point that empathy is correlated in the cognitive system by mirror neurons. The research undertaken by Rizzolatti's group [24] posits that mirror neurons form the biological basis of compassion and thereby of affective empathic experience, in the fields of motor, sensorimotor, language and emotions. For our purposes, it is the aesthetic experience which activates the mirror neurons, thus generating "motion, emotion and empathy" upon the representation that the work of art proposes, and also with the action of the artist performing/creating the artwork [15]. Berthoz [5] shows how the theory of mirror neurons proves that there is an intimate resonance between perception and action, and emphasizes how the brain implements the laws of physics which serve as anticipatory mechanisms of action.

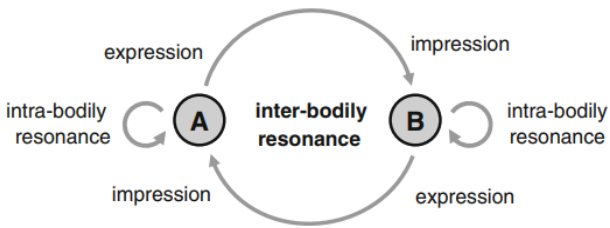

**Figure 1: The extended body from Fuchs et al. [16]**

## 1.3  The question of the resonance

In physics, the resonance phenomenon is defined as the property of a system or a body to vibrate with increasing or decreasing amplitudes at certain frequencies of excitation, generating harmonics with other systems or bodies. Berthoz also expresses that the brain does not passively receive visual information; rather, it solves ambiguities, predicts and anticipates, and also analyzes specific properties of the environment which correspond to specific tasks such as the need for face recognition, emotional body expressions, shapes, or movements of living creatures.

From a body perspective, Fuchs [17] proposes to conceive of the living being both as a "lived or subjective body" and as a "living or objective one," and the brain as a resonance organ which is mediating the circular interactions within the body, and the sensorimotor interactions between the body and the environment. Scioli et al. define the capacity for sharing deeper emotional states as "limbic resonance" [25]. For instance, relatives who reverberate with love, strength, and safety, either verbally or nonverbally, can engender a sense of connectedness, power, and security in each other. In mammals, it refers to the capacity for empathy and nonverbal connection, like a symphony of mutual exchange and internal adaptation [22]. Scioli et al. [25] underline that more than 75% of communication is nonverbal. When individuals mature, their ways of relating continue to evolve and expand, developing more complex "internal models" of connecting with others, called "emotional resonance" (figure 1). The research shows how this enactive approach treats conceiving of minds as fundamentally embodied and co-dependent, rather than as disembodied and independent from each other.

## 1.4 Virtual Reality as an experience medium

To assess the question of resonance, we choose a creative medium which is able to cover its range of complexity from intra-body and inter-body levels, to the relationship with the environment. To do so, we embark on applying the concept of resonance at different levels in the context of active and creative participation in installations conceived around the use of Virtual Reality (VR). A VR installation offers an elevated level of immersive and interactive experience due to the fact that the user's eyes and ears are isolated from his/her actual surroundings, and allows the engagement of the body through actions and expressivity. *Videoplace*, the famous pioneer artwork from Myron Krueger [20] in the 1980s, was conceived as an art medium which allowed the exploration of new ways for people to play with the virtual mirror of their body and objects, and even together from a distance. Since then, research in the field of VR, such as Slater [19] and Blanke's [8] current laboratories, acknowledge its increasing potential as a means of studying the sense of immersion, presence, self-location, agency, embodiment and body self-consciousness.

Our purpose is to observe the effects of improvisation and letting go applied to the field of arts and movement, and hypothesize that mobilizing body creativity and invoking playfulness might be sources of resonances. To do so, we propose the analysis of two examples of digital artworks based on VR and involving interaction between the user and the installation.

## 2  ANALYSIS OF ARTWORKS REGARDING RESONANCE

The two creations "InterACTE" and "Eve, dance is an unplaceable place" (later named "EVE") use VR as a medium, and involve playful body movement, dance, and the presence of an observing audience. Based on material produced by the authors and interviews, we propose to analyze the premises, the setup, and the reactions of the spectators in the two installations. Through these works, we have been able to draw some conclusions on the potentiality that both artworks show towards experiences that provoke resonance in the participating spectator and in the observing audience, identifying similar reactions.

## 2.1  InterACTE

Interactive Installation, 2015, by Dimitrios Batras, Judith Guez, Jean-François Jégo and Marie-Hélène Tramus.

*2.1.1 Artists' statement.* The interactive installation InterACTE allows improvisation using gestures with a virtual humanoid performer. It was inspired both by traditional shadow theater and by the Pierrot character from *Commedia dell'Arte* [2]. The installation was initially intended for use in theater training, but was then presented to a wider audience in different exhibitions and festivals. The virtual performer is able to improvise using a finite state machine to switch between three main behaviors: the virtual actor can mime in real-time the spectator's arm movements; it can also propose expressive or linguistic gestures from a motion capture database; and it can generate new arm gestures using a generative algorithm described by the authors in this article [2]. The body movements of the spectators which are tracked with a Kinect camera allow the triggering of the main behaviors.

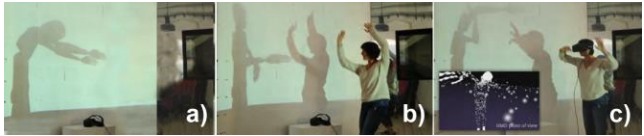

**Figure 2: InterACTE: a) virtual shadow waiting for a spectator, b) improvisation with the shadow of the virtual performer, c) immersive improvisation in the 3D virtual world using the VR headset and view from the VR headset**

*2.1.2 Artistic medium and setup.* The artwork was the result of an iterative process to create an expressive interactive virtual agent. The application design involves researchers who are linguists, digital artists and professional actors in theater. At first, in order to understand what a virtual gesture is, Batras et al. [2] explain that they designed a genetic algorithm that is able to generate purely new gestures. The fitness function is based on a database of motion captures of professional actors. Several experiments of improvisation were conducted with actors through computer-mediated interactions where one actor was playing the role of a virtual performer and the other one was playing the spectator. With this first setup, they could only communicate through feedback from a screen displaying the Kinect camera skeleton which has limited joints and no face tracking. Indeed, they mostly use gestures and it appears they start to imitate each other, taking turns, iterating each imitation with variation or totally new gestures. These experiments helped in creating an interactive platform where a virtual performer can perform an expressive improvisation enriched with the databases of generated or captured gestures. This allows the characterization of specific dynamics of gestures, recurrent behaviors and gestural patterns and the duration of each different behavior.

*2.1.3 Expected Scenario.* The performance is structured in three parts: an introduction where the spectator can simply look at the virtual actor's shadow which is playing a "waiting" (idle) animation captured on a mime actor (figure 2a). Then, the first contact with the virtual performer's shadow is triggered when the spectator gets closer, and the Kinect camera detects the participant, and then the virtual performer starts to imitate (figure 2b). The participant then can start to improvise with his/her own shadow together with the virtual character's shadow. In the third part, the participant is invited to continue playing with the virtual performer in an immersive 360 degrees virtual environment made of particles of light, using a VR headset (figure 2c). The virtual performer is placed at the exact same place which created the virtual shadow in the real world (figure 3). Thus, the spectator is immersed "alone" with the virtual performer without seeing the audience (figure 2c), whereas the audience is still able to see in the video projection the two shadows interacting.

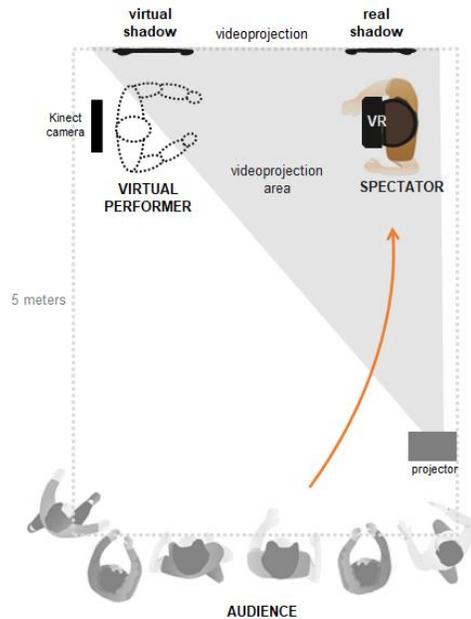

**Figure 3: Floor plan of the installation InterACTE showing a real spectator facing the virtual performer and the audience**

*2.1.4 Experience by spectators and feedback.* We present here the different experiences and feedback identified throughout several communications and interviews [2,28] produced by the authors of the artwork which we group into three categories. The gestural imitation of the virtual actor creates an immediate link with the spectator. Imitation is not a simple mirror, because the tracking is imperfect and a delay was created on purpose. Therefore, it can give the impression that the virtual actor follows the gestures of the spectator with hesitation. The creators reported that no participants needed specific instructions to start playing with the virtual performer nor needed help to understand the rules [2]. This was the case in many different cultural contexts: the installation was presented both in art exhibitions and conferences in Austria, France, Greece and Taiwan, to local and international participants. While other people were waiting to participate, they could watch the "play" and see how previous spectators used the system, react or improvise.

The authors reported that the spectators seemed to test the virtual character and tried to understand how it works and reacts. Some tried to touch its virtual body or to control it, especially during the imitation behavior. They were also surprised when the virtual performer proposed a new gesture (generated from its database or generating one on-the-fly using the genetic algorithm). The authors noted that some participants imitated the virtual actor reenacting its gestures and adapting to its degrees of freedom. For instance, they tested the reaction of the virtual performer when the imitation behavior was active: they tried to bend over, to jump, and to move the fingers whereas the virtual actor couldn't perform these actions. They also noted that spectators often adapted to the possibilities of the virtual actor performing wider and slower gestures or the body tracking fidelity. Some tried to hug the virtual performer which is actually designed at human scale and at reach of the spectator. Some participants even tried to find a certain harmony performing simpler gestures or sometimes finding other ways to communicate with the aim to create a "micro-relation" or their own stories.

The creators noted during Part 3 that when the spectators put on the VR headset, they got in front of the virtual performer but got no more visual representation of their own physical body. Naturally, they referred to the virtual body of the virtual performer especially when it was imitating. On the other hand, when the virtual actor was performing a gesture different from imitation, it looked more autonomous and the spectators could no longer identify their own gestures during this time. Also, since in this part the spectators were wearing the VR headset, they could not see the audience anymore and the authors reported more "letting go" from the spectators.

*2.1.5 Analysis.* From the feedback, we propose to analyze the identified key elements that we have grouped into main concepts:

• **Imitation and engagement** - The artwork was inspired by real improvisation experiments where real actors started to improvise, imitating each other. Since the scenario begins with imitation, it leads to a sort of simplicity and intuitiveness and in this case worked successfully in European and Asian cultures where it was presented. When the virtual performer proposes a new gesture different from imitation, this rupture tends to revive the dialogue by causing surprise, giving the illusion of a certain autonomy of the virtual actor. The spectator sometimes becomes influenced by the gesture of the virtual actor in a reflexive loop that produces, through an adaptive and reciprocal mimicry, a harmonious dialogue between the two actors.

Regarding the way the next participants were waiting to experience the artwork, they naturally seemed to share the rules of the game. This phenomenon evokes a "gestural contagion" and also reminds one of the "Chinese whispers" game between participants which in this case is based on gestures.

• **Degrees of freedom and playfulness** - A linguistic study conducted by Tramus et al. [28] showed that the virtual performer is not able to perform pronosupination, just like a toddler. In a certain way, the spectator is displaying childlike behavior when interacting with the virtual performer. And thus, a gestuality close to that of a toddler would generate empathy and invite to explore or to create other ways to interact, like the way we play with children.

• **Referentials** - In the VR part, the spectators have to deal with the absence of the vision of their own body referentials. So, they have to refer to the body of the virtual performer especially during the imitation behavior. According to a hypothesis from neuroscience researchers [27], this behavior refers to the state of empathy which is defined here as an experiential resonance to the other, a direct understanding of the others' emotional or mental state. When the virtual performer switches to behavior different from imitation, the spectators can no longer identify their own body and this situation gets closer to a state of sympathy which can be described as the act or capacity of entering into the feelings or interests of another. This installation is probably creating oscillations between empathy and sympathy.

Finally, regarding the VR headset itself, since it hides the view from the real world, it helps to give the spectators the feeling of being alone with the virtual performer in the virtual environment. So, the setup itself would indeed engage "letting go" more.

## 2.2 Eve, dance is an unplaceable place

Interactive performance, 2018, by Margherita Bergamo Meneghini and Daniel González.

*2.2.1 Artists' statement.* EVE, produced and launched in 2018 by Compagnie Voix and Omnipresenz, with further developments and showings in 2019, is a living installation combining Embodied Virtual Reality (EVR) stories to a live contemporary dance choreography, in interaction with a spectator and an external audience [3]. The authors created the artwork in the framework of a vision of dance as a cultural factor, an expressive element which can be reborn in the most intimate part of each person, helping to understand the other and the unconventional, and resolving personal traumas and social conflicts [18]. The project shows how the application of EVR exponentially multiplies the dance experience, the consciousness of a danced body/dancing body, and the possibility of expressing oneself with a language of movement [3,6]. Training spatialization, changing of points of view and creativity is the central goal of this artwork, viewed as a gateway to empathy.

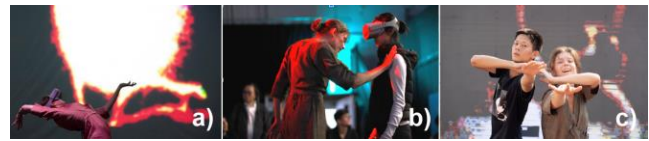

**Figure 4: EVE performance: a) real dancer introducing the story, b) spectator using the VR headset, c) spectator and dancer improvise together**

*2.2.2 Artistic medium and setup.* On stage, a professional dancer represents a ritual inspired by *Pachamama* (Mother Earth) bringing spectators to unexpected places, where they can virtually enter the body of a dancer, Eve, and where dance is a revolutionary language. During the body ownership illusionary experience (EVR), filmed characters get close to the participant and are part of the virtual environment, while generating a sense of presence [19]. On their end, the participant experiences the feeling of being the center of the narrative, and recognizes the new body as their own, tending to follow movements and acclimatizing to the new environment. An observing audience is witness to a ritual that includes the participation of the dancer and one spectator (figure 5): the virtual story is visible in the VR headset and through simultaneous video projection. The sound creates a specific atmosphere and soundscape, and connects several choreographies (the real dancer in live performance, the filmed dancers in the VR), driving the spectator to improvise movements following the dancer(s).

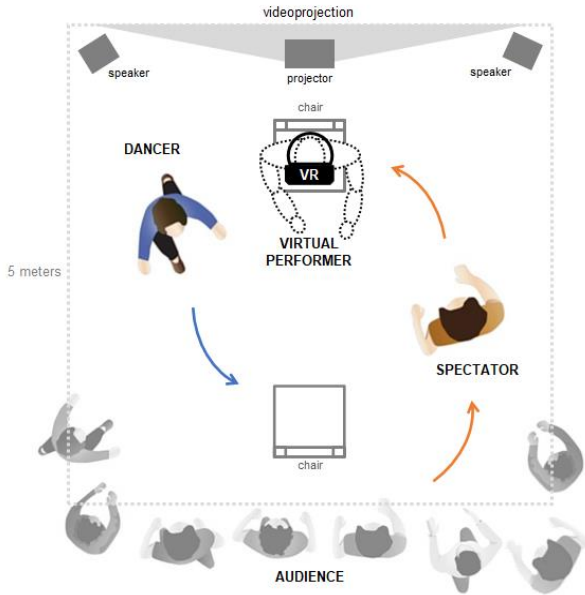

**Figure 5: Floor plan of the installation EVE showing a real spectator playing with a real dancer, the virtual and the audience**

*2.2.3 Expected Scenario.* EVE is structured in three parts: in the first part, a dancer wearing a VR headset introduces the participant-spectator to an abstract atmosphere, shown with the choreography and with an external flat screen projection (figure 4a). In the second part, the participant-spectator is welcomed by the dancer to change position in the installation setup, and to wear the same VR headset, embodying the avatar Eve in a story filmed in 360 degrees, and interacting with both the filmed dancers and the real dancer, through "embodiment" actions (figure 4b). Meanwhile the same 360-degree film is projected on a flat screen. In the third part, the participant-spectator and the dancer, both without any VR device, dance together alternating proposals on imitation and improvisation (figure 4c).

*2.2.4 Experience by spectators and feedback.* The authors kept track of the spectators' experiences at the end of all performances, avoiding any pre-established scheme and evaluating by asking for written feedback with pen and paper [3,4]. Through these notes, we were able to clearly detect the sense of feeling different from the normal; an opening towards the outside and towards the unknown; a superior connection; an emotion that comes from the experience of the moving body [3].

Participants follow the given rules to imitate the virtual avatar, Eve, guided by a real dancer. Most of them integrated their body well into the sequence of proposed movements. Moreover, they observed their body parts (such as arms, hands, legs) moving in concert with the virtual avatar's, and expressed having a conscious feeling of them. In general, the "embodiment" actions generated remarkable reactions due to the experiential sensation of touch, smell and taste, which empowered the predisposition to imitation, during and after the VR.

Some observing spectators spontaneously started dancing, and responded to the interaction with the dancer(s). Some others attended the performance many times, guided by curiosity and a sense of remote participation. Some participating spectators claimed to have completely forgotten reality, having integrated the action together with the virtual characters, and in some cases felt a great emotion manifested at the end of the experience.

After experiencing the whole progression of the experience, in the third part, participants are invited to express themselves dancing together with the real dancers. Some participants declared that this was the best part, and that the progression helped them to find their own creativity and to let go, and dance more.

*2.2.5 Analysis.* From the feedback, we have identified the key elements of EVE, as well as similar elements of InterACTE, that we have grouped into main concepts:

• **Imitation** - The goal of this project was to observe the effects of the imitation of a virtual body with a real body in movement, in order to enhance body expression and emotions through movement, and eventually generate empathy. The spectator recognized the new body as their own and tended toward spontaneously following the proposed movements: the predisposition to imitate confirms the inner simulation provided by mirror neurons [24]. In the last part of the performance, after the VR experience, the spectator reintegrated movement coordination with their real body, engaging with the rest of the real audience, and the real venue, eventually imitating the real dancer.

• **Engagement and presence** - EVR can perform as a communication interface [26]. The combined staging of the live and virtual performance opens up a wide variety of singular experiences of engagement and presence: spectators are surrounded by real and virtual dancers who are very close to them and a part of their environment, while generating a sense of presence also in VR, observing the impacts of human proximity in the virtual environment.

• **Creativity** - When the spectator is dancing, both in VR and in reality, spatialization and imagination are awakened, defining one's psychology and physicality, training individual ability to create, therefore to react to changes and find solutions.

## 3 DISCUSSION REGARDING RESONANCE

Based on the analysis of the two artworks, we identify different strategies to elicit improvisation and letting go regarding resonance. Here we propose to detail and analyze them following three perspectives. First, Resonance between Spectator and Artwork. Second, Resonance of the Spectator him/herself. Third, Resonance between Spectator / Artwork and Audience.

### 3.1 Resonance between Spectator and Artwork

*3.1.1 Identified concepts.* In InterACTE and EVE, the relationship between the spectator and the installation is based on imitation interspersed with moments of improvisation, where both the user and the installation (real dancer or virtual character) propose movements: the real dancer in EVE acts in a similar way to the virtual character in InterACTE.

• **The role of curiosity** - Curiosity is the first step to becoming involved in an unknown activity, and then starting toward an experiential path of knowledge. This curiosity can lead to engagement. VR is experienced through the use of headsets without which it is not possible to perceive the experience as it was conceived. This simple restriction generates curiosity and intrigues spectators. New media, digital art, and among them VR, are attractive expressive means of generating curiosity and wonder [13].

• **The role of playfulness** - Playfulness is a positive state for approaching learning through experience. It can awaken curiosity, and pave the way to a predisposition for integrating personal changes. Play with interactive digital art such as VR is nowadays more and more popular, and it can be driven in the direction of integrating body movements, self-expression and creativity.

• **The role of imitation** - The key element to start the improvisation is imitation. It can be focused on a body part position, or speed, or in the way the spectator has to follow a gesture. Imitation triggers the first sensorimotor improvisation. Imitation is the beginning of any learning process, a practice for understanding the other. Imitation of a body in motion requires presence, attention, concentration and eye contact. The act of imitating an example, with physical contact (in the case of EVE) or without physical contact (in the case of InterACTE), allows increasing the transmission of content and the retention of information.

• **The role of improvisation** - To improvise is to invent: increasing ability in body movement improvisation, based on previous experience and potentiality, overcomes self-induced limitations, in front of what appears to be the known/safe and the unknown/insecure. An individual must learn through experience how to generate and detect the appropriate perceptual information, in order to develop agency, prospectivity and flexibility [1].

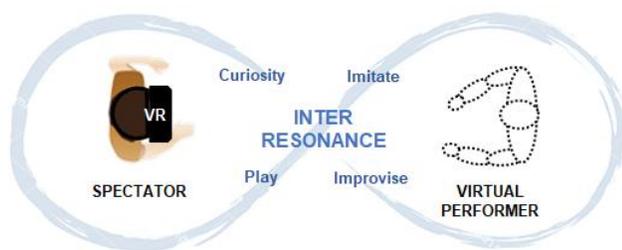

**Figure 6: Inter Resonance model**

*3.1.2 Inter Resonance .* In the framework of interactive artworks, which require the spectator to interact, play or perform, we study the possibility of the emergence of resonance. While the spectator is experiencing something new at several levels of action-perception, in real and virtual environments, s/he is playing with body movement, learning through experience, creatively proposing new movements.

## 3.2 Resonance of the Spectator him/herself

*3.2.1 Identified concepts.* Both in InterACTE and EVE, the participating spectator, once engaged in the interaction of the artwork, experiences and manifests a gradual integration with him/herself: the proposal originated by an external element - the artwork - passes to and inside the spectator, who generates responses and content.

• **The role of creativity** - Once a proposed practice is integrated, a personal practice can be developed using imagination, improvisation and personal skills. To launch the expression of creativity, to feed it and to maintain an open state of mind, we use the interactive artwork as a trigger.

• **The role of embodiment** - In both artworks, the spectator has to deal with his/her new body, which can be a shadow, an avatar or the video of someone else's body. The acceptance of this virtual body translates into engagement and sense of presence.

• **The role of being present** - The achievement of a state of "being present" of the body in motion, through play, imitation, improvisation and creation, takes us to the idea of a resonant communication from the body, the emergence of resonance in the bodily experience of an individual, towards the environment and others. The training of a mental-physical state through awareness can manifest a resonant body for communication, an extended body after an extended mind [21]. Moreover, the experience of a detachment from the real body during the virtual experience, allows to reflect on the virtual presence: the spectator experiences his/her personal presence in VR, and the presence of the virtual characters around him/her.

• **The role of letting go** - In VR, the headset isolates the user on an audiovisual level, and allows abstracting them from their "self" for a certain time. The user is "in play mode" in another reality far from their own. In the case of the two analyzed artworks, the VR environments propose movement, dancing, and practicing improvisation, themselves metaphors of an alternative way of expression. By "letting go," we're essentially referring to the body overcoming attached habits, physical and mental boundaries. Concurrently, this can be seen as a moving metaphor for pushing beyond personal limits in time and space. Through the experience of letting go, the body allows for reaching a deeper integration with the sensorimotor system, emotional state and thus with the surrounding environment.

*3.2.2 Inner Resonance.* During and after the interaction with the artworks, spectators experience a sense of emotional resonance. During the time that they get to know another environment, interact with virtual characters, and actively and creatively use their bodies, something changes within them. The artworks do not require functionality movement, they are a proposition to play with the act of moving and connecting between body parts. Spectators feel their body, create, let go, and awaken a sense of being present. The sense of emotional resonance which is created remains after the end of the experience, and according to given feedback, promotes personal psychophysical openness towards a state of empathy. Based on the illustration of the extended body from Fuchs et al. [17] we propose a model which presents the oscillations during the experience between internal motivation eliciting expression and letting go, and external environment producing impression while being present.

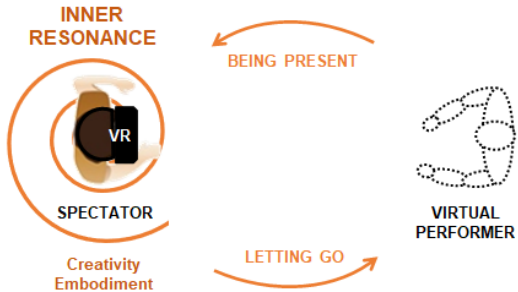

**Figure 7: Inner Resonance model**

## 3.3 Resonance between Spectator/Artwork and Audience

*3.3.1 Identified concepts.* A common feature in InterACTE and EVE is the presence of an observing audience. The spectator can experience the artwork without the functional need of an audience near them; but a solitary experience creates a restricted environment which is incomplete. Both artworks are designed to be experienced in a social environment where the gaze and the emotional interventions of the audience (noise, words, laughter) take importance: in this sense, the audience participates as well to the experiences. Firstly, the installations take place where the audience can casually pass by, so as to attract their curiosity. Secondly, the playfulness of the interactions gives rise to a certain attention, up to the moment when the audience wonders what the participant might be experiencing, and put themselves in his/her place, stimulating a sense of empathy. In the case of EVE, the professional dancer seeks visual and physical contact with much of the audience, to stimulate the idea that the audience is part of the choreographic experience.

• **The role of curiosity** - As detailed in part 3.1.1, curiosity is also involved when the audience is in contact with the artwork itself or in contact with a spectator interacting/playing with the artwork. It may trigger engagement to participate in the interactive experience.

• **The role of motor contagion** - Some instructions or a few mediators are needed to inform the spectator on how to start, and previous performances serve as examples to understand and convey interest: the inter-bodily experience (Inter Resonance) enriches the intra-bodily experience (Inner Resonance). Motor resonance is known to yield the kind of "mutual understanding" affecting the ability of humans to interact with others [7].

• **The role of engagement** - The artwork is a performance itself, where anyone is free to choose the level of engagement, as an observing or a participating spectator.

• **The role of gathering** - On the basis of Froese and Fuchs [16], we can recognize a shared bodily sensation and feeling which have the effect of lowering the psychological boundaries between the self and the group, and enhancing a sense of community and identity. As an example, during performances of EVE, some observing spectators spontaneously start dancing.

*3.3.2 Collective Resonance.* The path to a Collective Resonance starts in the artwork itself, which is the trigger for the experience, and awakens curiosity in the spectators. Once the spectator has chosen his/her role (as a participant or a viewer), s/he embarks on a journey of interaction with the artwork. The Inter Resonance is born within the interaction between the spectator and the artwork, generating feelings, memories and changes on different levels based on sensibility and background, developed in crescendo. This can provoke an increase of being present (impression), letting go (expression) and finally, an Inner Resonance in the spectator. The inner and Inter Resonance s can be reflected externally and reach the observing audience as ripples creating another level: the Collective Resonance. A connection emerges among all the elements of the installation and a sense of community simultaneously. Here, the real and/or virtual performers play a role in the installation as part of this community, and can finally share roles with the spectator, who then becomes the spectator-performer. The Collective Resonance takes place because each organism of the collectivity spontaneously regulates its coupling with the environment to create continuity in the generation of his/her identity [10]. It's therefore a commitment of the artwork (and the artists behind it) to choose the message to resonate to the collectivity.

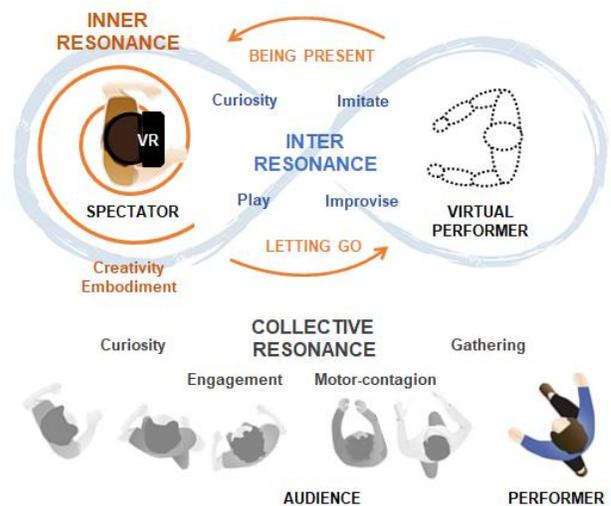

**Figure 8: Inner, Inter & Collective Resonances model**

## 4 CONCLUSION AND PERSPECTIVES

The two interactive artworks, and behind them the goal of their authors, are examples of how to use interactive digital art to awaken communicative potential and movement creativity. We analyzed the premises, the setup, and the feedback of the spectators in the two installations. We propose a model following three different perspectives of resonance (figure 8): 1. Inter Resonance between Spectator and Artwork, which involves curiosity, imitation, playfulness and improvisation. 2. Inner Resonance of Spectator him/herself, where embodiment and creativity contribute to perceiving the sense of being present and letting go. 3. Collective Resonance between Spectator/Artwork and Audience, which is stimulated by curiosity, and triggers motor contagion, engagement and gathering.

When we move, create and express ourselves with the body, we activate our sensorimotor system, our emotional state, our relationships with the environment and with others. We stimulate and reflect on body parts, the way they influence each other, in the whole-body ecosystem. Body parts resonate among them, within the environment and within others, as a wide resonance chamber made by the body-brain-environment ecosystem [17]. The engine of this resonance could be movement, understanding that movement is already generated in the ascertainment of the presence and the intention, and develops through improvisation when creativity is let go. The substantial fact lies in seeing the body, or bodies, as an active and proactive participant in the vital movement of this resonance chamber.

The core of these perspectives, which make us ask "why" we detect the existence and effects of resonance, is perhaps now sliding towards a more complex area: we propose the idea that resonance ripples are enhanced between perception and action as waves in human bodies through presence and movement. Humans, like other animals, continuously generate body movements, but only conscious movements are connected with our state of being present. When we improvise and experiment with creative body movements, we create movement beyond the conventional, we awaken the "being present" and rouse intuition, facilitating adaptability towards change, renewal and freer expression. The analyzed examples come to light in a social moment which reveals a struggle of humanity to understand and support one another to achieve peace, having an effect on interpersonal, local or global scales.

On another level, the same question about the emergence of resonance reveals the current emergency of open-minded communication in contemporary societies. Perhaps it is up to artists to trigger the urgency of resonance: to find ways to elicit it through artistic practices such as improvisation and letting go, using contemporary media such as interactive digital art, in order to actualize communicative and collective possibilities which are already present, but still need to be empowered.

## ACKNOWLEDGMENTS

The authors wish to thank all the artists, the scientists and the participants who contribute to the study. We also wish to thank Mark J. Lee for proofreading. "InterACTE" was funded by the Labex Arts-H2H (ANR-10-LABX-80-01) program from the University Paris 8. "Eve, dance is an unplaceable place" was produced with the support of Tangente Danse and Festival de Nouveau cinéma de Montréal.